\documentclass[12pt,prd,tightenlines,nofootinbib,showpacs,showkeys,setspace]{revtex4-2}
\usepackage{bm}
\usepackage{graphics}
\usepackage{rotating}
\usepackage{epsfig}
\usepackage{amsmath}
\usepackage{amsfonts}
\usepackage{setspace}
\usepackage{amssymb}
\newcommand{\be}{\begin{equation}}
\newcommand{\ee}{\end{equation}}
\newcommand{\bdis}{\begin{displaymath}}
\newcommand{\edis}{\end{displaymath}}
\newcommand{\bga}{\begin{equation}\begin{gathered}}
\newcommand{\ega}{\end{gathered}\end{equation}}
\newcommand{\mathsym}[1]{{}}
\newcommand{\unicode}[1]{{}}

\begin{document}
\title{Higgs boson decay to the pair of S- and P-wave $B_c$ mesons}
\author{\firstname{R.~N.} \surname{Faustov}}
\affiliation{Institute of Cybernetics and Informatics in Education, FRC CSC RAS, Moscow, Russia}
\author{\firstname{F.~A.} \surname{Martynenko}}
\affiliation{Samara University, Samara, Russia}
\author{\firstname{A.~P.} \surname{Martynenko}}
\affiliation{Samara University, Samara, Russia}

\begin{abstract}
We investigate the rare production of the pair of S- and P-wave $B_c$-meson in the Higgs boson 
decay in the perturbative Standard Model and relativistic quark model.
Relativistic amplitudes and decay widths are constructed with the account
of the relative motion of heavy quarks forming $B_c$ mesons.
When constructing the Higgs boson decay amplitudes, the method of projection operators 
on the S- and P-wave states of quarks is used.
Relativistic corrections are expressed in terms of special relativistic parameters and 
are calculated numerically in the quark model.
The dependence of the decay widths of the Higgs boson on various sources of relativistic 
corrections is investigated.
\end{abstract}



\maketitle

\normalsize
\section{Introduction}

The study of properties of the Higgs boson and its decays began immediately 
after its discovery by the ATLAS and CMS experiments at the LHC \cite{atlas,cms}.
At present, a detailed study of the Higgs sector is the important direction in particle 
physics.
The large mass of the Higgs boson, as well as significant coupling constants with gauge 
bosons and quarks, provide a variety of decay channels \cite{keung,vysotsky,bodwin,luchinsky,qiao,niu,kataev}.
The analysis of various decays is based on a set of proton-proton collision data 
collected with detectors at the CERN Large Hadron Collider.
Main decay channels of the Higgs boson are connected with the production
of photons, W, Z bosons and heavy quarks.
Investigation of various mechanisms of heavy quarkonia production in Higgs boson decays 
is of obvious interest in connection with testing various approaches to describing 
the production of heavy quarks and their bound states.
Among the decay processes of the Higgs boson, one can single out a rare exclusive process 
in which a pair of $B_c$ mesons is formed in different S- and P-states. For a reliable description 
of the amplitudes of such processes, it is important to have a consistent theory of the production 
of a pair of bound states of quarks and antiquarks, in which the effects of the relative motion 
of heavy quarks are strictly taken into account.
Such processes are also interesting in the sense that a pair of $B_c$ mesons is produced here, 
which, in contrast to charmonium and bottomonium, have been studied to a much lesser 
extent experimentally.
Bound states of two different heavy quarks
$(\bar b c)$ with open beauty and charm stand out in a significant way among heavy quarkonia
since its decay mechanism differs significantly from the decay mechanism of charmonium or bottomonium. 
There is a hope that the processes of pair production of $B_c$ mesons in various states can 
be investigated experimentally from the decay products of $B_c$ mesons.
In this work, we continue the study of relativistic effects during the production of a pair 
of $B_c$ mesons in Higgs boson decays \cite{apm2021}, including in the field of study not only 
S-wave, but also P-wave $B_c$ mesons.
We study the relativistic corrections both in the production amplitude of two quarks 
and antiquarks, and in the transformation law of the wave functions of $B_c$ mesons during 
the transition from the rest frame to the moving reference frame.
As  known from our previous studies the corrections of these types significantly  contribute to  the production cross section of a pair of heavy quarkonia \cite{apm1,apm2,apm4}.

Our approach to the study of Higgs boson decays includes a perturbative stage, when two free 
quarks and two antiquarks are created, and a nonperturbative stage, in which the formation 
of quarks bound states occurs.
The nonperturbative part of the meson production process is described within the framework 
of the relativistic quark model.
This mechanism of production of a pair of $B_c$ mesons, which we call quark, is further considered 
in detail in the following sections.
Within this approach, we can systematically take into account relativistic effects at all stages 
of the decay process, including the formation of $B_c$ mesons due to the strong interaction 
of quarks.
Taking into account relativistic effects in the production of heavy quarkonia in various reactions 
is very important for achieving good accuracy in calculating the observed quantities \cite{qwg2011}.

One of the first works devoted to the pair production of quarkonia in Higgs boson decays was done in the nonrelativistic approximation in \cite{keung}. 
The production of single quarkonia in the H decay was investigated in \cite{vysotsky,bodwin} 
with the account of relativistic corrections and one-loop corrections.
In the work \cite{luchinsky}, various channels of the Higgs boson decay into pairs of heavy quarkonia were studied, including $H\to ZZ$, $H\to WW$.
Single $B_c$ meson and double heavy baryon production rates in Higgs boson decays were calculated 
in the nonrelativistic QCD framework in \cite{qiao,niu}.
The observations of the Higgs boson decays into a pair $\gamma\gamma$, $WW$, $ZZ$, $b\bar b$ and
$\tau\tau$ have been reported in \cite{aad1,aad2,cms2,cms3}.
The first experimental searches for decays of the Higgs boson into a pair of $J/\Psi$ and 
$\Upsilon$ mesons were performed in \cite{cms1}.
If the bound states of the same heavy quarks and antiquarks have been studied experimentally 
well enough, the bound states of different heavy quarks were observed much less frequently \cite{bll2019}. In fact, $B_c$ mesons are known only for the $1S$ and $2S$ states. Therefore, 
the study of various mechanisms for the production  of S-wave and P-wave $B_c$ mesons is 
of obvious interest, which is connected with the study of their properties.

\section{General formalism}

\begin{figure}[htbp]
\centering
\includegraphics[scale=0.3]{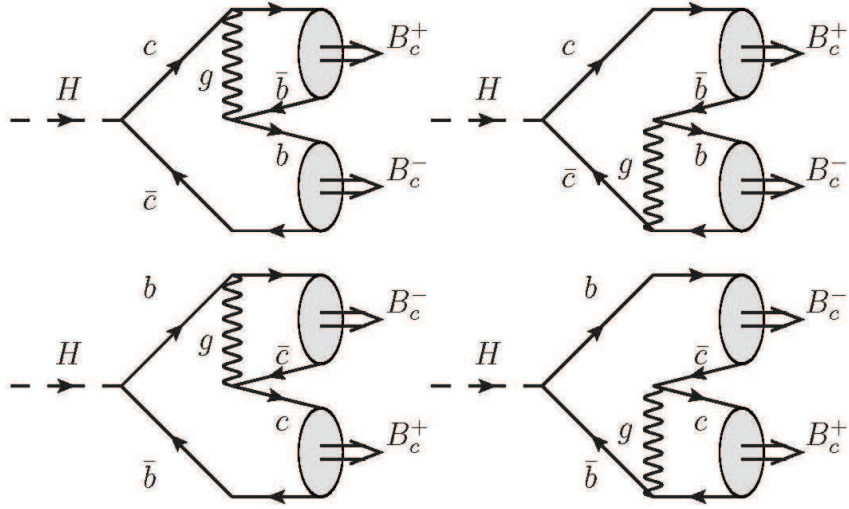}
\caption{The quark mechanism for the production of a pair of $B_c$ mesons in the decay 
of the Higgs boson. Dashed line shows the Higgs boson
and wavy line corresponds to the gluon.}
\label{fig1}
\end{figure}

Four quark production amplitudes of the $B_c$ meson pair in leading order of the QCD coupling 
constant $\alpha_s$ are presented in Fig.~\ref{fig1}.
We investigate the production channel of a pair of $B_c$ mesons connected with the initial production 
of a pair of heavy quarks $b$ or $c$ in the Higgs boson decay.
There are two stages of $B_c$ meson production process. At the first stage, which is described
by the perturbative Standard Model, the Higgs boson transforms into a heavy quark-antiquark pair. Then
the heavy quark or antiquark emits a virtual gluon which produces another heavy quark-antiquark pair.
At the second stage, heavy quarks and antiquarks combine with some probability into bound states.

Four-momenta of heavy quarks and antiquarks can be expressed in terms of relative and total 
four momenta as follows:
\begin{equation}
\label{eq:pq}
p_1=\eta_{1}P+p,~p_2=\eta_{2}P-p,~(p\cdot P)=0,~\eta_{i}=
\frac{M_1^2\pm m_1^2\mp m_2^2}{2M_1^2},
\end{equation}
\begin{displaymath}
q_1=\rho_{1}Q+q,~q_2=\rho_{2}Q-q,~(q\cdot Q)=0,~\rho_{i}=
\frac{M_2^2\pm m_1^2\mp m_2^2}{2M_2^2},
\end{displaymath}
where $M_1$ and $M_2$ are the masses of $B_c^+$ and $B_c^-$ mesons
consisting of $\bar bc$ and $\bar cb$.
$m_{1,2}$ are the masses of $c$ and $b$ quarks. 
Neglecting the effects of particle coupling, we obtain $\rho_1\approx\eta_1\approx m_1/(m_1+m_2)$,
$\rho_2\approx\eta_2\approx m_2/(m_1+m_2)$.
$P$, $Q$ are the total four-momenta of mesons $B_c^+$ and $B_c^{-}$, relative quark four-momenta
$p=L_P(0,{\bf p})$ and
$q=L_P(0,{\bf q})$ are obtained from the rest frame four-momenta $(0,{\bf p})$ and $(0,{\bf q})$ 
by the Lorentz transformation to the system moving with the momenta $P$ and $Q$.
The index $i=1,2$ corresponds to plus and minus signs in \eqref{eq:pq}.
Heavy quarks $c$, $b$ and antiquarks $\bar c$, $ \bar b$ are outside 
the mass shell in the intermediate state:
$p_{1,2}^2=\eta_{1,2}^2P^2-{\bf p}^2=\eta_{1,2}^2M_1^2-{\bf p}^2\not= m_{1,2}^2$,
so that $p_1^2-m_1^2=p_2^2-m_2^2$.

In this work we study the production of S-wave and P-wave $B_c$ mesons. 
It is convenient to begin the construction of the pair production 
amplitudes with a description of the production of S-wave $B_c$ states.
Let consider the production amplitude of two pseudoscalar and two vector $B_c$ mesons
setting $M_1=M_2=M_{B_c}$. Initially it can be written as a
convolution of perturbative production amplitude of free quarks and antiquarks and 
the quasipotential wave
functions of $B_c$ mesons moving with four-momenta P and Q. Using then the transformation 
law of the bound state wave functions from the rest frame to the
moving one with four-momenta $P$ and $Q$ we can present the meson production amplitude in the form
\cite{apm1,apm2,apm3}:
\begin{equation}
\label{eq:amp}
{\cal M}(P,Q)=-i(\sqrt{2}G_F)^{\frac{1}{2}}\frac{2\pi}{3}M_{B_c}
\int\frac{d{\bf p}}{(2\pi)^3}\int\frac{d{\bf q}}{(2\pi)^3}\times
\end{equation}
\begin{displaymath}
\times Tr\left\{\Psi^{\cal P,V}_{B_c}(p,P)\Gamma_1^{\nu}(p,q,P,Q)\Psi^{\cal P,V}_{B_c}(q,Q)
\gamma_\nu+
\Psi^{\cal P,V}_{B_c}(q,Q)\Gamma_2^{\nu}(p,q,P,Q)\Psi^{\cal P,V}_{B_c}(p,P)
\gamma_\nu\right\},
\end{displaymath}
where a superscript ${\cal P}$ indicates a pseudoscalar $B_c$ meson, a superscript ${\cal V}$ 
indicates a vector $B_c$ meson, $G_F$ is the Fermi constant. $\Gamma_{1,2}$ 
are the vertex functions defined below.
The permutation of subscripts $b$ and $c$ in the wave functions indicates corresponding 
permutation in the projection operators (see below Eqs.\eqref{eq:amp1}-\eqref{eq:amp2}).
The method for producing the amplitudes in the form \eqref{eq:amp} is described in detail in our previous studies
\cite{apm3,apm4,apm5}.
The transition of free quark-antiquark pair to meson bound states is described in our approach by specific wave functions.
Relativistic wave functions of pseudoscalar and vector $B_c$ mesons accounting for the transformation from the
rest frame to the moving one with four momenta $P$, and $Q$ are
\begin{eqnarray}
\label{eq:amp1}
\Psi^{\cal P}_{B_c}(p,P)&=&\frac{\Psi^0_{B_c}({\bf p})}{
\sqrt{\frac{\epsilon_1(p)}{m_1}\frac{(\epsilon_1(p)+m_1)}{2m_1}
\frac{\epsilon_2(p)}{m_2}\frac{(\epsilon_2(p)+m_2)}{2m_2}}}
\left[\frac{\hat v_1-1}{2}+\hat
v_1\frac{{\bf p}^2}{2m_2(\epsilon_2(p)+ m_2)}-\frac{\hat{p}}{2m_2}\right]\cr
&&\times\gamma_5(1+\hat v_1) \left[\frac{\hat
v_1+1}{2}+\hat v_1\frac{{\bf p}^2}{2m_1(\epsilon_1(p)+
m_1)}+\frac{\hat{p}}{2m_1}\right],
\end{eqnarray}
\begin{eqnarray}
\label{eq:amp2}
\Psi^{\cal V}_{B^\ast_c}(q,Q)&=&\frac{\Psi^0_{B^\ast_c}({\bf q})}
{\sqrt{\frac{\epsilon_1(q)}{m_1}\frac{(\epsilon_1(q)+m_1)}{2m_1}
\frac{\epsilon_2(q)}{m_2}\frac{(\epsilon_2(q)+m_2)}{2m_2}}}
\left[\frac{\hat v_2-1}{2}+\hat v_2\frac{{\bf q}^2}{2m_1(\epsilon_1(q)+
m_1)}+\frac{\hat{q}}{2m_1}\right]\cr &&\times\hat{\varepsilon}(Q,S_z)(1+\hat v_2) 
\left[\frac{\hat v_2+1}{2}+\hat v_2\frac{{\bf q}^2}{2m_2(\epsilon_2(q)+ m_2)}-\frac{\hat{q}}{2m_2}\right],
\end{eqnarray}
where the symbol hat denotes convolution of four-vector with the Dirac gamma matrices,
$v_1=P/M_{B_c}$, $v_2=Q/M_{B_c}$;
$\varepsilon^\lambda(Q,S_z)$ is the polarization vector of the $B^{\ast-}_c(1^-)$ meson,
relativistic quark energies $\epsilon_{1,2}(p)=\sqrt{p^2+m_{1,2}^2}$.
In~\eqref{eq:amp1} and \eqref{eq:amp2} we have complicated factor depending on relative
momenta ${\bf p}$, ${\bf q}$ including the bound state wave 
function in the rest frame $\Psi^0_{B_c}({\bf p}) $.
The production amplitude \eqref{eq:amp} contains the
integration over the quark relative momenta ${\bf p}$ and ${\bf q}$.
The result of integration essentially depends on the wave functions of bound states 
of heavy quarks.
The color part of the meson wave function in the amplitude~\eqref{eq:amp} is taken as 
$\delta_{ij}/\sqrt{3}$ (color indexes $i, j, k=1, 2, 3$).
Relativistic wave functions in~\eqref{eq:amp1} and \eqref{eq:amp2} are equal to the product 
of wave functions in the rest frame
$\Psi^0_{B_c}({\bf p})$ and spin projection operators that are
accurate at all orders in $|{\bf p}|/m$ and $|{\bf q}|/m$. An expression of spin projector 
in different form for $(c\bar c)$ system was obtained in \cite{bodwin2002} where spin projectors are
written in terms of heavy quark momenta $p_{1,2}$ lying on the mass shell.
Our derivation of relations~\eqref{eq:amp1} and \eqref{eq:amp2} accounts for the transformation
law of the bound state wave functions from the rest frame to the
moving one with four momenta $P$ and $Q$. This transformation law was discussed in the Bethe-Salpeter
approach in \cite{brodsky} and in quasipotential method in \cite{faustov}.

\subsection{Production of a pair of $B_c$ mesons in S-states}

When constructing the decay amplitudes with the production of a pair of S-wave $B_c$ mesons, 
we introduce projection operators $\hat\Pi^{\cal P,V}$ for states with total spin $S=0,~1$
of the following form:
\begin{equation}
\label{eq:uu}
\hat\Pi^{\cal P}=[v_2(0)\bar u_1(0)]_{S=0}=\gamma_5\frac{1+\gamma^0}{2\sqrt{2}},~~~
\hat\Pi^{\cal V}=[v_2(0)\bar u_1(0)]_{S=1}=\hat\varepsilon\frac{1+\gamma^0}{2\sqrt{2}}.
\end{equation}

After that, the total amplitude of the Higgs boson decay in the leading order in strong 
coupling constant $\alpha_s$ can be represented in the form:
\begin{equation}
\label{eq:A1}
{\mathcal M}=\frac{4\pi}{3}M_{B_c}
\int\!\frac{d\mathbf{p}}{(2\pi)^3}\int\!\frac{d\mathbf{q}}{(2\pi)^3}
\mathrm{Tr}\bigl\{\mathcal T_{12}+\mathcal T_{34}\bigr\},
\end{equation}
\begin{equation}
\label{t12}
{\mathcal T_{12}}=\Gamma_c\alpha_{b}\Psi^{\cal P,V}_{B_c}(p,P)
\left[\frac{\hat p_1-\hat r+m_1}{(r-p_1)^2-m_1^2}\,\gamma_\mu+\gamma_\mu\,\frac{\hat r-\hat q_1+m_1}{(r-q_1)^2-m_1^2}\right]D^{\mu\nu}(k_2)
\Psi^{\cal P,V}_{B_c}(q,Q)\gamma_\nu,
\end{equation}
\begin{equation}
\label{t34}
{\mathcal T_{34}}=\Gamma_b\alpha_{c}\Psi^{\cal P,V}_{B_c}(q,Q)
\left[\frac{\hat p_2-\hat r+m_2}{(r-p_2)^2-m_2^2}\,\gamma_\mu+
\gamma_\mu\frac{\hat r-\hat q_2+m_2}{(r-q_2)^2-m_2^2}\,\right]D^{\mu\nu}(k_1)
\Psi^{\cal P,V}_{B_c}(p,P)
\gamma_\nu,
\end{equation}
where subscripts 12, 34 denote the contributions of amplitudes 1 and 2, 3 and 4 in Fig.~\ref{fig1},
$\alpha_{c,b}=\alpha_s\left(m_{1,2}^2M_H^2/(m_1+m_2)^2\Lambda^2\right)$.
$\Gamma_c=m_1(\sqrt{2} G_F)^{\frac{1}{2}}$, $\Gamma_b=m_2(\sqrt{2} G_F)^{\frac{1}{2}}$.
The trace calculation in \eqref{eq:A1}
leads to amplitudes ${\cal M}_{PP}$ and ${\cal M}_{VV}$ presented in~\eqref{eq:amp11}-\eqref{eq:amp22}.
Four-momentum of Higgs boson squared $r^2=M_H^2=(P+Q)^2=2M_{B_c}^2+2PQ$,
the gluon four-momenta are $k_1=p_1+q_1$, $k_2=p_2+q_2$.
Relative momenta $p$, $q$ of heavy quarks enter in the gluon propagators $D_{\mu\nu}(k_{1,2})$
and quark propagators as well as in relativistic wave functions~\eqref{eq:amp1} and \eqref{eq:amp2}.
Accounting for the small ratio of relative quark momenta $p$ and $q$ to the mass of the Higgs boson $M_H$, we can simplify the inverse denominators of quark and gluon propagators as follows:
\begin{equation}
\label{den1}
\frac{1}{(p_1+q_1)^2}=\frac{1}{\eta_1^2 M_H^2},~~~
\frac{1}{(p_2+q_2)^2}=\frac{1}{\eta_2^2 M_H^2},
\end{equation}
\begin{equation}
\label{den2}
\frac{1}{(r-q_1)^2-m_1^2}=\frac{1}{\eta_2 M_H^2},~~~
\frac{1}{(-r-p_1)^2-m_1^2}=\frac{1}{\eta_2 M_H^2},
\end{equation}
\begin{equation}
\label{den3}
\frac{1}{(r-p_2)^2-m_1^2}=\frac{1}{\eta_1 M_H^2},~~~
\frac{1}{(-r-q_2)^2-m_1^2}=\frac{1}{\eta_1 M_H^2}.
\end{equation}
In the case of $B_c$ meson production of the same mass $\rho_1=\eta_1$, $\rho_2=\eta_2$.
The formulas \eqref{den1}-\eqref{den3} mean that we completely neglect corrections of the form
$|{\bf p}|/M_H$, $|{\bf q}|/M_H$. 
At the same time, we keep in the amplitudes \eqref{t12}, \eqref{t34} the second-order correction 
for small ratios $|{\bf p}|/m_{1,2}$, $|{\bf q}|/m_{1,2}$
relative to the leading order result.
As we take relativistic factors in the denominator of the amplitudes \eqref{eq:amp1} and \eqref{eq:amp2} unchanged,
the momentum integrals are convergent. Calculating the trace in obtained expression
in the package FORM \cite{form}, we find relativistic amplitudes of the $B_c$ meson pairs production
in the form:
\begin{equation}
\label{eq:amp11}
{\cal M_{PP}}=\frac{32\pi}{3M_H^4}(\sqrt{2}G_F)^{\frac{1}{2}}M^2_{B_c}
\left[\frac{\alpha_{b}r_1}{\eta_2^3}F_{1P}+\frac{\alpha_{c}r_2}{\eta_1^3}F_{2P}\right]
|\tilde\Psi_{\cal P}(0)|^2,
\end{equation}
\begin{equation}
\label{eq:amp22}
{\cal M_{VV}}=\frac{32\pi}{3M_H^4}(\sqrt{2}G_F)^{\frac{1}{2}}M^2_{B_c} \varepsilon_1^{\lambda}\varepsilon_2^{\sigma}
\left[\frac{\alpha_{b}r_1}{\eta_2^3}F_{1V}^{\lambda\sigma}+\frac{\alpha_{c}r_2}
{\eta_1^3}F_{2V}^{\lambda\sigma}\right]|\tilde\Psi_{\cal V}(0)|^2,
\end{equation}
where $\varepsilon_1^\lambda$, $\varepsilon_2^\sigma$ are the polarization vectors of spin 1 
$B_c$ mesons, $r_1=m_1/M_{B_c}$, $r_2=m_2/M_{B_c}$, the parameter $r_3=\frac{M_H}{M_{B_c}}$.
The decay widths of the Higgs boson into a pair of pseudoscalar and vector $B_c$ mesons are 
determined by the following expressions \cite{apm2021}:
\begin{equation}
\label{gapp}
\Gamma_{\cal{PP}}=\frac{512\sqrt{2}\pi G_F |\tilde\Psi_{\cal P}(0)|^4 \sqrt{\frac{r_3^2}{4}-1}}
{9M_H^3 r_3^7}
\left[\frac{\alpha_{b}r_1}{\eta_2^3}F_{1P}+\frac{\alpha_{c}r_2}{\eta_1^3}F_{2P}\right]^2,
\end{equation}
\begin{equation}
\label{gapp1}
F_{1P}=-r_1-\eta_1+\frac{3}{2}r_3^2-\frac{1}{2}r_1 r_3^2-\frac{1}{2}\eta_1 r_3^2+\omega_{01}
(-12r_2+2r_2r_3^2)+
\end{equation}
\begin{displaymath}
\omega_{10}(2r_1+r_1r_3^2)+\omega_{10}\omega_{01}(6r_2-2r_1-\frac{3}{2}r_3^2-
r_3^2r_2-r_3^2r_1),
\end{displaymath}
\begin{equation}
\label{gavv}
\Gamma_{\cal{VV}}=\frac{512\sqrt{2}\pi G_F  |\tilde\Psi_{\cal V}(0)|^4 
\sqrt{\frac{r_3^2}{4}-1}}{9M_H^3 r_3^7}\sum_{\lambda,\sigma}
\vert\varepsilon_1^\lambda\varepsilon_2^\sigma
\left[\frac{\alpha_{b}r_1}{\eta_2^3}F^{\lambda\sigma}_{1V}+\frac{\alpha_{c}r_2}{\eta_1^3}
F^{\lambda\sigma}_{2V}\right]\vert^2,
\end{equation}
\begin{equation}
\label{gavv1}
F^{\alpha\beta}_{1V}=g_1 v_1^\alpha v_2^\beta+g_2g^{\alpha\beta},~~~
F^{\alpha\beta}_{2V}=\tilde g_1 v_1^\alpha v_2^\beta+\tilde g_2g^{\alpha\beta},
\end{equation}
\begin{displaymath}
g_1=-1+\frac{1}{9}\omega_{10}\omega_{01},~~~g_2=-r_1-\eta_1+\frac{1}{2}r_3^2-\frac{4}{3}r_2\omega_{01}+
2r_1\omega_{10}+\omega_{10}\omega_{01}(\frac{2}{3}r_2-\frac{2}{9}-\frac{1}{18}r_3^2).
\end{displaymath}

The functions $F_{2P}$, $F_{2V}^{\alpha\beta}$ entering in the production amplitude 
\eqref{gapp}, \eqref{gavv} 
can be obtained from $F_{1P}$, $F_{1V}^{\alpha\beta}$
changing $r_2\leftrightarrow r_1$, $m_1\leftrightarrow m_2$ and $\omega_{ij}\to\omega_{ji}$, 
$\eta_{1,2}\leftrightarrow \rho_{2,1}$.

General expressions for the decay rates \eqref{gapp}, \eqref{gavv} contain numerous parameters.
One part of the parameters, such as quark masses, the masses of $B_c$ mesons are determined 
within the framework of quark models as a result of calculating the observed quantities. 
The parameters of quark models are found from the condition of the best agreement with 
experimental data.
Another part of the relativistic parameters can also be found in the quark model as 
a result of calculating integrals with wave functions of quark bound states in the 
momentum representation.
The corresponding calculation results are discussed in subsection C.

\subsection{Production of a pair of $B_c$ mesons in S- and P-states}

The production of $B_c$ mesons in the P-state has its own specific features, which we 
discuss in this section. We further consider the production of one S-wave and one 
P-wave $B_c$ meson. For definiteness, let us further consider the construction of the pair 
production amplitude in the states $^3S_1$ and $^3P_J$, where ${\bf J}={\bf S}+{\bf L}$,
$J=0,~1,~2$ is the total momentum of the pair of quark and antiquark with spin $S=1$ and
orbital momentum $L=1$. Initially, we introduce two spin polarization vectors, since both 
quark-antiquark pairs are in states with spin 1:
\begin{equation}
\label{eq:AA1}
{\mathcal M}({^3S_1}+{^3P_J})=\frac{4\pi}{3}M_{B_c}
\int\!\frac{d\mathbf{p}}{(2\pi)^3}\int\!\frac{d\mathbf{q}}{(2\pi)^3}
\mathrm{Tr}\bigl\{\mathcal M_{12}+\mathcal M_{34}\bigr\},
\end{equation}
\begin{equation}
\label{tt12}
{\mathcal M_{12}}=\frac{\Gamma_c\alpha_{b}}{\tilde r_2^3 M_H^4}\Psi^{\cal V}_{B_c}(p,P)
\left[(\hat p_1-\hat r+m_1)\gamma_\mu+\gamma_\mu(\hat r-\hat q_1+m_1)\right]
\Psi^{\cal V}_{B_c}(q,Q)\gamma_\mu,
\end{equation}
\begin{equation}
\label{tt34}
{\mathcal M_{34}}=\frac{\Gamma_b\alpha_{c}}{\tilde r_1^3 M_H^4}\Psi^{\cal V}_{B_c}(q,Q)
\left[(\hat p_2-\hat r+m_2)\gamma_\mu+
\gamma_\mu(\hat r-\hat q_2+m_2)\right]
\Psi^{\cal V}_{B_c}(p,P)\gamma_\mu,
\end{equation}
where we neglect bound state corrections in factors $\eta_{1,2}$, $\rho_{1,2}$ 
from the propagator denominators
setting $\eta_{1,2}\approx \tilde r_{1,2}=m_{1,2}/(m_1+m_2)$, 
$\rho_{1,2}\approx \tilde r_{1,2}=m_{1,2}/(m_1+m_2)$.
After calculating the trace in \eqref{eq:AA1} and expanding the amplitudes ${\mathcal M_{12}}$
and ${\mathcal M_{34}}$ in powers of $p$ to the second 
order and $q$ to the third order, we introduce the polarization vector of the orbital motion
as follows:
\begin{equation}
\label{vec_pol}
\int\frac{d{\bf q}}{(2\pi)^3}q_\mu \psi_{LL_z}({\bf q})=-i\varepsilon_\mu(L_Z)\sqrt{\frac{3}{4\pi}}
R'_P(0),
\end{equation}
where $R'_P(0)$ is the derivative of the radial wave function at zero.
Further, when adding the spin and orbital angular momentum, we can distinguish individual states 
with the total angular momentum $J=0,~1,~2$:
\begin{equation}
\label{j012}
\Psi_{\alpha\beta}(J,J_z)=\sum_{L_z,S_z}<1,L_z;1,S_z|J,J_z>\varepsilon_\alpha(v_2,L_z)
\varepsilon_\beta(v_2,S_z)=
\end{equation}
\begin{displaymath}
=\begin{cases}
\frac{1}{\sqrt{3}}(g_{\alpha\beta}-v_{2\alpha}v_{2\beta}),~J=0,\\
\frac{i}{\sqrt{2}}\varepsilon_{\alpha\beta\mu\nu}v_2^\mu\varepsilon^\nu(v_2,J_z),~J=1,\\
\varepsilon_{\alpha\beta}(v_2,J_z),~J=2.\\
\end{cases}
\end{displaymath}
It is convenient to separate the relativistic corrections of the required order and add 
the moments in accordance with \eqref{j012} in the package Form \cite{form}.
Omitting some factors in \eqref{eq:AA1}, including the value of radial wave
function at zero $R_S(0)$ and $R'_P(0)$,
the production amplitudes for states ${^3S_1}+{^3P_J}$ can be presented as
follows:
\begin{equation}
\label{j0}
{\mathcal N_1}({^3S_1}+{^3P_0})=\frac{16M}{3M_H^4}(\sqrt{2}\pi G_F)^{\frac{1}{2}}
\left[\frac{\alpha_{b}}{\tilde r_2^3}\left(f_1+g_1\right)+\frac{\alpha_{c}}{\tilde r_1^3}
\left(\tilde f_1+
\tilde g_1\right)\right]\left(v_2^\alpha \varepsilon_\alpha(v_1,S_z)\right),
\end{equation}
\begin{equation}
\label{j1}
{\mathcal N_2}({^3S_1}+{^3P_1})=\frac{16M}{\sqrt{6}M_H^4}(\sqrt{2}\pi G_F)^{\frac{1}{2}}
\bigl[\frac{\alpha_{b}}{\tilde r_2^3}\bigl(f_2+g_2\bigr)+\frac{\alpha_{c}}{\tilde r_1^3}
\bigl(\tilde f_2+
\tilde g_2\bigr)\bigr]\varepsilon_{\mu\nu\alpha\beta}v_1^\mu v_2^\nu\varepsilon^\beta(v_1,S_z)
\varepsilon^\alpha(v_2,J_z),
\end{equation}
where functions $f_i$, $\tilde f_i$ (i=1,2,3,4,5) denote a purely nonrelativistic contribution, 
and $g_i$, $\tilde g_i$ (i=1,2,3,4,5) denote relativistic corrections, $M=m_1+m_2$.
An explicit form of these functions is presented in Appendix~A.
Note also that the amplitude ${\mathcal M}({^3S_1}+{^3P_2})$
vanishes taking into account the properties of the symmetric tensor $\varepsilon_{\alpha\beta}$
describing the state with J=2:
$ \varepsilon_{\alpha\beta}g_{\alpha\beta}=0 $, $ \varepsilon_{\alpha\beta}v_2^\alpha=0$.

Other amplitudes for the production of S- and P-states are constructed in a similar way. 
For completeness, we also present here their general form. 
In the case of the final state of a pair of $B_c$ mesons, one of the mesons has a spin equal 
to zero (${^3S_1}$ state). 
The orbital angular momentum of the quark-antiquark pair gives the total angular momentum $J = 1$.
The initial expression for the amplitude is determined by formula \eqref{eq:A1} with one wave function 
$\Psi^{\cal V}_{B_c}(p,P)$ and another wave function $\Psi^{\cal P}_{B_c}(q,Q)$.
The antisymmetric tensor that appears when calculating the trace in \eqref{eq:A1} ultimately 
determines the general structure of the amplitude, which can be represented as follows:
\begin{equation}
\label{jl1}
{\mathcal N_3}({^3S_1}+{^1P_1})=\frac{16M}{\sqrt{3}M_H^4}(\sqrt{2}\pi G_F)^{\frac{1}{2}}
\bigl[\frac{\alpha_{b}}{\tilde r_2^3}\bigl(f_3+g_3\bigr)+\frac{\alpha_{c}}{\tilde r_1^3}
\bigl(\tilde f_3+
\tilde g_3\bigr)\bigr]\varepsilon_{\mu\nu\alpha\beta}v_1^\mu v_2^\nu\varepsilon^\beta(v_1,S_z)
\varepsilon^\alpha(v_2,J_z).
\end{equation}

So far, we have discussed the production amplitudes in which one of the $B_c$ mesons is 
in the state ${^3S_1}$. There are also amplitudes in which this meson is produced in a state 
$^1S_0$ with spin $S=0$.
If the second meson is also produced in a singlet spin state, the amplitude of this process 
takes the form:
\begin{equation}
\label{jsl1}
{\mathcal N_4}({^1S_0}+{^1P_1})=\frac{16M}{\sqrt{3}M_H^4}(\sqrt{2}\pi G_F)^{\frac{1}{2}}
\bigl[\frac{\alpha_{b}}{\tilde r_2^3}\bigl(f_4+g_4\bigr)+\frac{\alpha_{c}}{\tilde r_1^3}
\bigl(\tilde f_4+
\tilde g_4\bigr)\bigr](v_1^\alpha\varepsilon_\alpha(v_2,J_z)).
\end{equation}

Finally, the last nonzero amplitude corresponds to the process ${^1S_0}+{^3P_1}$
and has the form:
\begin{equation}
\label{jsl11}
{\mathcal N_5}({^1S_0}+{^3P_1})=\frac{16M}{\sqrt{6}M_H^4}(\sqrt{2}\pi G_F)^{\frac{1}{2}}
\bigl[\frac{\alpha_{b}}{\tilde r_2^3}\bigl(f_5+g_5\bigr)+\frac{\alpha_{c}}{\tilde r_1^3}
\bigl(\tilde f_5+
\tilde g_5\bigr)\bigr](v_1^\alpha\varepsilon_\alpha(v_2,J_z)).
\end{equation}

Two other amplitudes connected with the production of states ${^1S_0}+{^3P_0}$
and ${^1S_0}+{^3P_2}$ vanish. In the case of state ${^1S_0}+{^3P_0}$ the production
amplitude is proportional to the convolution $\varepsilon_{\mu\nu\alpha\beta}v_1^\mu v_2^\nu
(g^{\alpha\beta}-v_2^\alpha v_2^\beta)\equiv 0$. For the state ${^1S_0}+{^3P_2}$
we also have a convolution of symmetric and antisymmetric tensors:
$\varepsilon_{\mu\nu\alpha\beta}v_1^\mu v_2^\nu\varepsilon^{\alpha\beta}\equiv 0$.
In both reactions, the conservation law of momentum and parity is not satisfied.

The decay widths of the Higgs boson into a pair of different S- and P-wave $B_c$ mesons 
with masses $M_1$ (S-wave meson) and $M_2$ (P-wave meson)
for all the states discussed above can be represented in the following form:
\begin{equation}
\label{gasp1}
\Gamma({^{1}S_0}+{^{3}P_1})=\frac{16\sqrt{2}|{\bf P}|G_FM^2}{3\pi r_3^4M_H^6M_1M_2}
|\tilde R_S(0)|^2|\tilde R'_P(0)|^2\left[\frac{1}{4}(r_3^2-r_4-\frac{1}{r_4})^2-1\right]\times
\end{equation}
\begin{displaymath}
\Biggl\{\frac{\alpha_b}{\tilde r_2^3}(f_1+g_1)+\frac{\alpha_c}{\tilde r_1^3}
(\tilde f_1+\tilde g_1)\Biggr\}^2,~~~|{\bf P}|=\frac{1}{2M_H}\sqrt{[M_H^2-(M_1-M_2)^2]
[M_H^2-(M_1+M_2)^2]},
\end{displaymath}
\begin{equation}
\label{gasp2}
\Gamma({^{1}S_0}+{^{1}P_1})=\frac{32\sqrt{2}|{\bf P}|G_FM^2}{3\pi r_3^4M_H^6M_1M_2}
|\tilde R_S(0)|^2|\tilde R'_P(0)|^2\left[\frac{1}{4}(r_3^2-r_4-\frac{1}{r_4})^2-1\right]\times
\end{equation}
\begin{displaymath}
\Biggl\{\frac{\alpha_b}{\tilde r_2^3}(f_2+g_2)+\frac{\alpha_c}{\tilde r_1^3}
(\tilde f_2+\tilde g_2)\Biggr\}^2,
\end{displaymath}
\begin{equation}
\label{gasp3}
\Gamma({^{3}S_1}+{^{3}P_0})=\frac{32\sqrt{2}|{\bf P}|G_FM^2}{9\pi r_3^4M_H^6M_1M_2}
|\tilde R_S(0)|^2|\tilde R'_P(0)|^2\left[\frac{1}{4}(r_3^2-r_4-\frac{1}{r_4})^2-1\right]\times
\end{equation}
\begin{displaymath}
\Biggl\{\frac{\alpha_b}{\tilde r_2^3}(f_3+g_3)+\frac{\alpha_c}{\tilde r_1^3}
(\tilde f_3+\tilde g_3)\Biggr\}^2,
\end{displaymath}
\begin{equation}
\label{gasp4}
\Gamma({^{3}S_1}+{^{3}P_1})=\frac{8\sqrt{2}|{\bf P}|G_FM^2}{3\pi r_3^4M_H^6M_1M_2}
|\tilde R_S(0)|^2|\tilde R'_P(0)|^2\left[\frac{1}{4}(r_3^2-r_4-\frac{1}{r_4})^2-1\right]\times
\end{equation}
\begin{displaymath}
\Biggl\{\frac{\alpha_b}{\tilde r_2^3}(f_4+g_4)+\frac{\alpha_c}{\tilde r_1^3}
(\tilde f_4+\tilde g_4)\Biggr\}^2,
\end{displaymath}
\begin{equation}
\label{gasp5}
\Gamma({^{3}S_1}+{^{1}P_1})=\frac{32\sqrt{2}|{\bf P}|G_FM^2}{3\pi r_3^4M_H^6M_1M_2}
|\tilde R_S(0)|^2|\tilde R'_P(0)|^2\left[\frac{1}{4}(r_3^2-r_4-\frac{1}{r_4})^2-1\right]\times
\end{equation}
\begin{displaymath}
\Biggl\{\frac{\alpha_b}{\tilde r_2^3}(f_5+g_5)+\frac{\alpha_c}{\tilde r_1^3}
(\tilde f_5+\tilde g_5)\Biggr\}^2.
\end{displaymath}

The results of the numerical calculation of the decay widths using these formulas are 
presented in the Table~\ref{tb2} and in the final section of the work.

\subsection{Relativistic parameters in the decay widths}

The amplitudes of the pair production of $B_c$ mesons in the decay of the Higgs boson 
are expressed in terms of functions $F_{i P}$,  $F_{i V}$, $f_i$, $g_i$, $\tilde f_i$,
$\tilde g_i$ that are presented in the form of an expansion in ${|\bf p}|/m_{1,2}$,
${|\bf q}|/m_{1,2}$ up to terms of the second order.
As a result of algebraic transformations, it turns out to be convenient to express 
relativistic corrections in terms of relativistic factors 
$C_{ij}=[(\epsilon_1(p)-m_1)/(\epsilon_1(p)+m_1)]^i
[(\epsilon_2(q)-m_2)/(\epsilon_2(q)+m_2)]^j$ with $i+j\leq 2$.
When these factors are integrated with the wave functions of bound states of quarks, 
$\omega^{P,V}_{nk}$ quantities arise, which represent a set of relativistic parameters 
of the theory. In the case of S-states $\omega^{P,V}_{nk}$ are determined by 
the momentum integrals $I_{nk}$ in the form:
\begin{equation}
\label{eq:intnk}
I_{nk}^{P,V}=\int_0^\infty p^2R^{P,V}(p)\sqrt{\frac{(\epsilon_1(p)+m_1)(\epsilon_2(p)+m_2)}
{2\epsilon_1(p)\cdot 2\epsilon_2(p)}}
\left(\frac{\epsilon_1(p)-m_1}{\epsilon_1(p)+m_1}\right)^n
\left(\frac{\epsilon_2(p)-m_2}{\epsilon_2(p)+m_2}\right)^k dp,
\end{equation}
\begin{equation}
\label{eq:parameter}
\omega^{P,V}_{10}=\frac{I^{P,V}_{10}}{I^{P,V}_{00}},~
\omega^{P,V}_{01}=\frac{I^{P,V}_{01}}{I^{P,V}_{00}},~
\omega^{P,V}_{\frac{1}{2}\frac{1}{2}}=
\frac{I^{P,V}_{\frac{1}{2}\frac{1}{2}}}{I^{P,V}_{00}},~
\omega^{P,V}_{20}=\frac{I^{P,V}_{20}}{I^{P,V}_{00}},~
\omega^{P,V}_{02}=\frac{I^{P,V}_{02}}{I^{P,V}_{00}},~\omega^{P,V}_{11}=
\frac{I^{P,V}_{11}}{I^{P,V}_{00}},
\end{equation}
\begin{equation}
\tilde R(0)=\frac{\sqrt{2}}{\sqrt{\pi}}\int_0^\infty \sqrt{\frac{(\epsilon_1(p)+m_1)
(\epsilon_2(p)+m_2)}{2\epsilon_1(p)\cdot 2\epsilon_2(p)}}p^2R(p)dp,
\end{equation}
where superscripts $P,V$ denote pseudoscalar and vector states.

In the case of P-states (L=1) we denote relativistic parameters $\tilde\omega_{nk}$. They are determined by the momentum integrals $J_{nk}$ in the form:
\begin{equation}
\label{eq:intnk1}
J_{nk}=\int_0^\infty q^3R(q)\sqrt{\frac{(\epsilon_1(q)+m_1)(\epsilon_2(q)+m_2)}
{2\epsilon_1(q)\cdot 2\epsilon_2(q)}}
\left(\frac{\epsilon_1(q)-m_1}{\epsilon_1(q)+m_1}\right)^n
\left(\frac{\epsilon_2(q)-m_2}{\epsilon_2(q)+m_2}\right)^k dq,
\end{equation}
\begin{equation}
\label{eq:parameter_1}
\tilde\omega_{10}=\frac{J_{10}}{J_{00}},~\tilde\omega_{01}=\frac{J_{01}}{J_{00}},~
\tilde\omega_{\frac{1}{2}\frac{1}{2}}=\frac{J_{\frac{1}{2}\frac{1}{2}}}{J_{00}},~
\tilde\omega_{20}=\frac{J_{20}}{J_{00}},~
\tilde\omega_{02}=\frac{J_{02}}{J_{00}},~\tilde\omega_{11}=\frac{J_{11}}{J_{00}},
\end{equation}
\begin{equation}
\tilde R'(0)=\frac{\sqrt{2}}{3\sqrt{\pi}}\int_0^\infty \sqrt{\frac{(\epsilon_1(q)+m_1)
(\epsilon_2(q)+m_2)}{2\epsilon_1(q)\cdot 2\epsilon_2(q)}}q^3R(q)dq.
\end{equation}

Another source of relativistic corrections is related with the Hamiltonian of the heavy quark 
bound states which allows to calculate 
the bound state wave functions of pseudoscalar, vector $B_c$ mesons (S-states) and 
wave functions for the P-states.
The exact form of the bound state wave functions $\Psi^0_{B_c}({\bf q})$
is important to obtain more reliable predictions for the decay widths.
In the nonrelativistic approximation the Higgs boson decay width with a production of a pair 
of $B_c$ mesons contains the fourth power of the nonrelativistic wave function at the origin 
for S-states or second power of derivative of the radial wave function at zero for P-states. 
The value of the decay width is very sensitive 
to small changes of $\Psi^0_{B_c}(0)$, $R_S(0)$, $R'_P(0)$. In the nonrelativistic QCD there 
exists corresponding problem of determining the magnitude of the
color-singlet matrix elements \cite{bbl}. To account for relativistic corrections to the meson 
wave functions we describe the dynamics of heavy quarks by the QCD generalization of
the standard Breit Hamiltonian in the center-of-mass reference frame 
\cite{repko1,pot1,capstick,godfrey,glko,godfrey1,lucha1995,rqm1,rqm2,rqm3}:
\begin{equation}
\label{eq:breit}
H=H_0+\Delta U_1+\Delta U_2,~~~H_0=\sqrt{{\bf
p}^2+m_1^2}+\sqrt{{\bf p}^2+m_2^2}-\frac{4\tilde\alpha_s}{3r}+(Ar+B),
\end{equation}
\begin{equation}
\label{eq:breit1}
\Delta U_1(r)=-\frac{\alpha_s^2}{3\pi r}\left[2\beta_0\ln(\mu
r)+a_1+2\gamma_E\beta_0
\right],~~a_1=\frac{31}{3}-\frac{10}{9}n_f,~~\beta_0=11-\frac{2}{3}n_f,
\end{equation}
\begin{equation}
\label{eq:breit2}
\Delta U_2(r)=-\frac{2\alpha_s}{3m_1m_2r}\left[{\bf p}^2+\frac{{\bf
r}({\bf r}{\bf p}){\bf p}}{r^2}\right]+\frac{2\pi
\alpha_s}{3}\left(\frac{1}{m_1^2}+\frac{1}{m_2^2}\right)\delta({\bf r})+
\frac{4\alpha_s}{3r^3}\left(\frac{1}{2m_1^2}+\frac{1}{m_1m_2}\right)({\bf S}_1{\bf L})+
\end{equation}
\begin{displaymath}
+\frac{4\alpha_s}{3r^3}\left(\frac{1}{2m_2^2}+\frac{1}{m_1m_2}\right)({\bf S}_2{\bf L})
+\frac{32\pi\alpha_s}{9m_1m_2}({\bf S}_1{\bf S}_2)\delta({\bf r})+
\frac{4\alpha_s}{3m_1m_2r^3}\left[\frac{3({\bf S}_1{\bf r})({\bf S}_2{\bf r})}{r^2}-
({\bf S}_1{\bf S}_2)\right]-
\end{displaymath}
\begin{displaymath}
-\frac{\alpha_s^2(m_1+m_2)}{m_1m_2r^2}\left[1-\frac{4m_1m_2}{9(m_1+m_2)^2}\right],
\end{displaymath}
where ${\bf L}=[{\bf r}\times{\bf p}]$, ${\bf S}_1$, ${\bf S}_2$ are spins of heavy quarks,
$n_f$ is the number of flavours, $\gamma_E\approx 0.577216$ is
the Euler constant. To improve an agreement of theoretical hyperfine splitting in $(\bar bc)$ mesons
with experimental data and other calculations in quark models we add to the standard Breit potential \eqref{eq:breit2}
the spin confining potential obtained in \cite{repko1,repko2,gupta,gupta1}:
\begin{equation}
\Delta V^{hfs}_{conf}(r)=
f_V\frac{A}{8r}\left\{\frac{1}{m_1^2}+\frac{1}{m_2^2}+\frac{16}{3m_1m_2}({\bf S}_1{\bf S}_2)+
\frac{4}{3m_1m_2}
\left[\frac{3({\bf S}_1{\bf r})({\bf S}_2{\bf r})}{r^2}-({\bf S}_1{\bf S}_2)\right]
\right\},
\end{equation}
where we take the parameter $f_V=0.9$. For the dependence of the
QCD coupling constant $\tilde\alpha_s(\mu^2)$ on the renormalization point
$\mu^2$ in the pure Coulomb term in~\eqref{eq:breit} we use the three-loop result 
\cite{kniehl1997}:
\begin{equation}
\label{26}
\tilde\alpha_s(\mu^2)=\frac{4\pi}{\beta_0L}-\frac{16\pi b_1\ln L}{(\beta_0 L)^2}+\frac{64\pi}{(\beta_0L)^3}
\left[b_1^2(\ln^2 L-\ln L-1)+b_2\right], \quad L=\ln(\mu^2/\Lambda^2).
\end{equation}
In other terms of the Hamiltonians~\eqref{eq:breit1} and \eqref{eq:breit2} we use
the leading order approximation for $\alpha_s$. The typical momentum transfer scale in a
quarkonium is of order of double reduced mass, so we set the renormalization scale 
$\mu=2m_1m_2/(m_1+m_2)$ and $\Lambda=0.168$ GeV.
The coefficients $b_i$ are written explicitly in \cite{kniehl1997}.
The parameters of the linear potential $A=0.18$ GeV$^2$ and $B=-0.16$ GeV have established 
values in quark models.

Starting with Hamiltonian \eqref{eq:breit}, we construct an effective quark model 
for the S- and P-states. 
The main details of this model are described in Appendix B and our previous work \cite{apm2}.
The numerical values of the relativistic parameters entering the decay widths~\eqref{gapp}, 
\eqref{gavv}, \eqref{gasp1}-\eqref{gasp5} are obtained by the numerical solution of the 
Schr\"odinger equation \cite{LS}. They are presented in Table~\ref{tb1}.
For comparison, we present in Table~\ref{tb1} the values of some parameters $\tilde\omega_{20}$,
$\tilde\omega_{02}$, $\tilde\omega_{11}$, that are omitted 
in analytical expressions, since they give corrections of order of $O({\bf q}^4)$, 
$O({\bf p}^4)$.

\begin{table}[h]
\caption{Numerical values of the relativistic parameters~\eqref{eq:parameter}, 
\eqref{eq:parameter_1}.}
\bigskip
\label{tb1}
\begin{ruledtabular}
\begin{tabular}{|c|c|c|c|c|c|c|c|c|}
 $n^{2S+1}L_J$ &$M_{B_c}$, &$\tilde R(0)$, & $\omega_{10},$ &$\omega_{01},$
 & $\omega_{\frac{1}{2}\frac{1}{2}},$ &$\omega_{20},$ &$\omega_{02},$ & $\omega_{11},$   \\
& GeV  &  $\tilde R'(0)$ & $\tilde\omega_{10}$   &  $\tilde\omega_{01}$  & 
$\tilde\omega_{\frac{1}{2}\frac{1}{2}}$  &  $\tilde\omega_{20}$ & $\tilde\omega_{02}$ &    
$\tilde\omega_{11}$     \\    \hline
$1^1S_0$ & 6.275 & 0.886 & 0.0728 & 0.0089&0.0254 & 0.0073  & 0.0001&0.0009  \\
  &  &   &   &   &  & &   &     \\  \hline
 $1^3S_1$  & 6.317 & 0.750 & 0.0703  & 0.0086 & 0.0245   & 0.0069  &  0.0001   
 &0.0009  \\
  &  &   &   &   &  & &   &    \\  \hline
$2^3P_2$ & 6.757  & 0.371   &0.1020 &0.0129 & 0.0362 & 0.0121 & 0.0002  & 0.0016  \\  \hline
$2^1P_1$ & 6.736  & 0.531   & 0.1035 &  0.0131 &  0.0368 & 0.0124   &0.0002 &  0.0016    \\  \hline
$2^3P_1$ & 6.726  &  0.319  &  0.0998  & 0.0126   &  0.0354   & 0.0116 & 0.0002 &  0.0015 \\  \hline
$2^3P_0$ &6.688   & 0.281 & 0.0981 &  0.0123  &  0.0347 &   0.0113  & 0.0002 &  0.0014  \\  \hline
\end{tabular}
\end{ruledtabular}
\end{table}

\begin{table}[h]
\caption{Numerical results for the decay widths in the nonrelativistic approximation
and with the account for relativistic corrections.}
\bigskip
\label{tb2}
\begin{ruledtabular}
\begin{tabular}{|c|c|c|}
Final state & Nonrelativistic decay width $\Gamma_{nr} \cdot 10^{14}$  & Relativistic decay width $\Gamma_{rel} \cdot 10^{14}$  \\
$B_{\bar bc}B_{b\bar c}$    &    in GeV      &  in GeV \\  \hline
$1^1S_0+1^1S_0$   & 125  &   45     \\  \hline
$1^3S_1+1^3S_1$   & 125  &   20    \\  \hline
$1^1S_0+2^3P_1$   & $0.32$   & $0.12$        \\  \hline
$1^1S_0+2^1P_1$   & $0.77$   & $0.59$        \\  \hline
$1^3S_1+2^3P_0$   & $9.81$  & $1.45$       \\  \hline
$1^3S_1+2^3P_1$   & $0.87$   & $0.16$        \\  \hline
$1^3S_1+2^1P_1$   & $0.16$   & $0.10$        \\  \hline
\end{tabular}
\end{ruledtabular}
\end{table}

\begin{figure}[htbp]
\centering
\includegraphics[scale=1.]{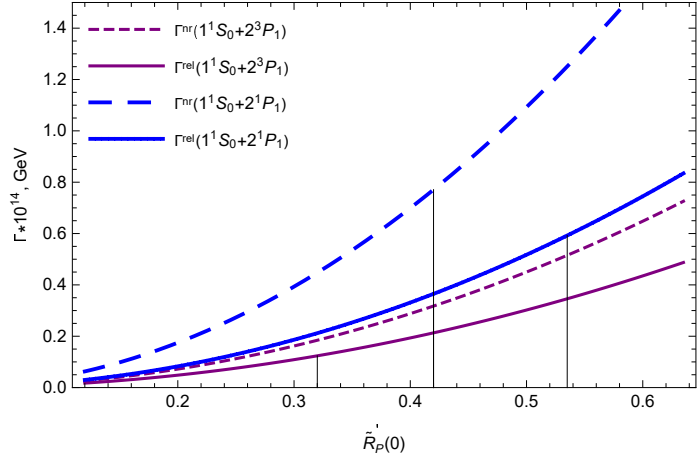}
\caption{The dependence of the $H$ boson decay width on the value 
$R'_P(0)$.}
\label{fig2}
\end{figure}

\begin{figure}[htbp]
\centering
\includegraphics[scale=1.]{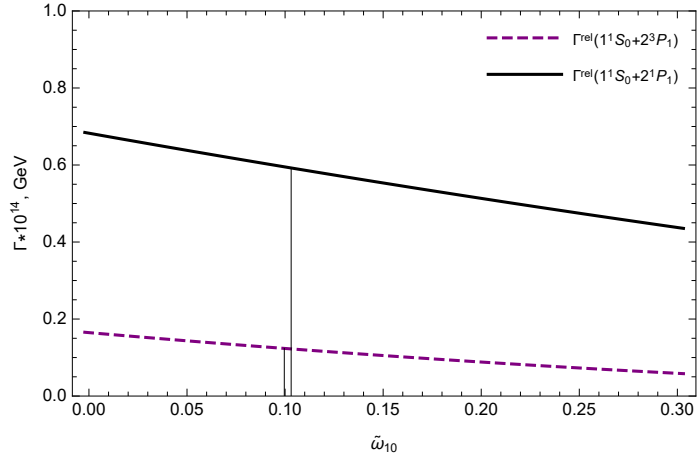}
\caption{The dependence of the $H$ boson decay width on the value 
of the largest relativistic parameter $\tilde\omega_{10}$.}
\label{fig3}
\end{figure}

\begin{figure}[htbp]
\centering
\includegraphics[scale=1.]{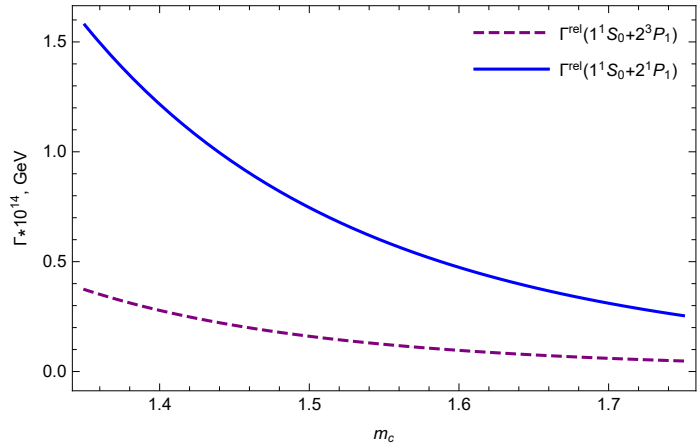}
\caption{The dependence of the $H$ boson decay width on the value of the c-quark mass.}
\label{fig4}
\end{figure}

\section{Numerical results and conclusion}

In this work, we study rare exclusive decays of the Higgs boson into a pair of 
$B_c$ mesons in different S- and P-states. The investigation of various decay channels 
of the Higgs boson acquired particular relevance after its discovery, although before that, 
numerous works were carried out to study various mechanisms and decay processes. 
The study of rare exclusive processes plays an important role in checking the
fundamentals of the theory and determining the values of basic parameters.
In our work, special 
attention is paid to relativistic effects connected with taking into account the relative motion 
of heavy quarks that form $B_c$ mesons at the end of the reaction. First, when constructing 
the Higgs boson decay amplitudes, we take into account the laws of transformation of the 
relativistic wave functions of mesons and various factors that depend on the relative momenta 
of the quarks. Then, taking into account the smallness of these relative momenta with respect 
to the mass of the Higgs boson, some simplifying transformations are made with the extraction 
of second-order corrections in relative momenta. Relativistic expressions are obtained for the 
widths of the Higgs boson decay into a pair of different S- and P-wave $B_c$ mesons, on the 
basis of which numerical estimates of these decay widths are made. The decay widths are calculated 
in the nonrelativistic approximation, and the obtained nonrelativistic results are compared with 
the results taking into account relativistic effects. We analyzed the dependence of the total 
decay widths on various sources of relativistic corrections both in the relativistic amplitude 
of the production of two heavy quarks and antiquarks, and in the operator of the interaction 
of quarks that compose $B_c$ mesons. The role of the main factors $R_S(0)$ and $R'_P(0)$, which affects 
the numerical results for decay widths, is highlighted. The total degree of these two factors 
is four, so even a 30 percent decrease in $R_S(0)$ and $R'_P(0)$ leads to a fourfold decrease 
in the decay widths.

The dependence of the Higgs boson decay width at the production of S- and P-wave 
$B_c$ mesons on the key parameters $R'_P(0)$, relativistic factor $\tilde\omega_{10}$ 
and the mass of c-quark $m_1$ is shown in Fig.~\ref{fig2}, Fig.~\ref{fig3}, Fig.~\ref{fig4}.
Our calculations clearly show that taking into account relativistic corrections in the 
interaction potential of quarks leads to a decrease in the values of $R'_P(0)$, and, as a consequence, 
to a decrease in the values of decay widths. When constructing these plots, a fixed value 
of the wave function at zero for S-states is used. Relativistic corrections in decay amplitudes 
are determined by $\omega_{ij}$, $\tilde\omega_{ij}$ factors. The dependence 
on the largest relativistic factor $\tilde\omega_{10}$ which describes relativistic effects 
for P-wave mesons is shown in Fig.~\ref{fig3}. At the 
same time, the value of $\omega_{10}$ for S-states was also chosen to be fixed. 
In this case, there is a decrease in decay widths (some of them are shown in Fig.~\ref{fig3}) 
not exceeding $10\%$. In a separate Fig.~\ref{fig4} the dependence of the calculation results
on the c-quark mass for several processes is presented.
The obtained numerical values of the decay widths are small in comparison with the 
total width of Higgs boson $\Gamma=3.2^{+2.8}_{-2.2}\cdot 10^{-3}$ GeV \cite{pdg}. Therefore, 
to observe rare decay processes with the formation of a pair of $B_c$ mesons, it is necessary 
to increase the luminosity of the LHC.
Possibly that rare decays of the Higgs boson could be investigated at other Higgs factories 
besides the LHC, 
such as the ILC (International Linear Collider) and CLIC (Compact Linear Collider)
\cite{clic}.

\begin{figure}[htbp]
\centering
\includegraphics[scale=0.4]{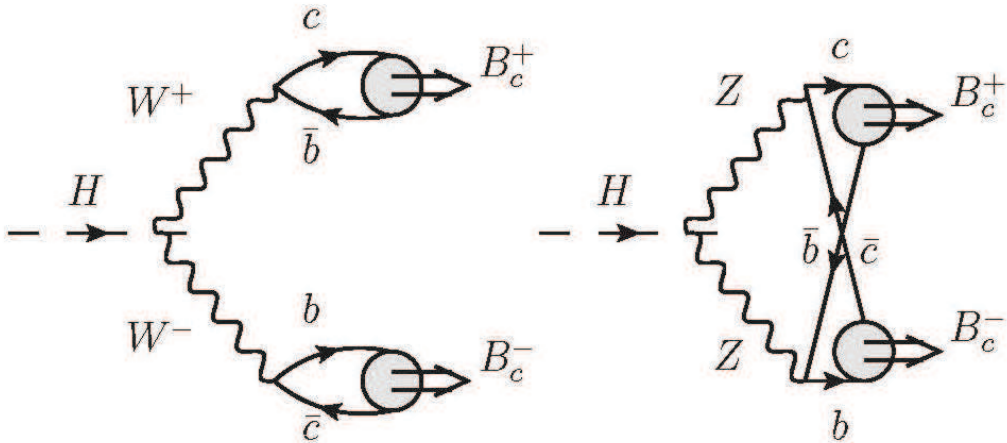}
\caption{Some amplitudes of the $Z,W$ boson production of a pair of $B_c$ mesons in the decay 
of the Higgs boson.}
\label{fig56}
\end{figure}

In this work, a purely quark mechanism for the production of a pair of B mesons in the decay of the 
H boson is investigated. There are other production mechanisms, which are determined, for example, 
by the initial decay of the H boson into a pair of ZZ, WW and by other couplings 
(see Fig.~\ref{fig56}).
We can obtain an estimate of such contributions using the formalism presented
in the previous sections. Considering for definiteness the decay of the Higgs boson into 
a pair of $ZZ$ with the subsequent production of a pair of pseudoscalar $B_c$ mesons, 
we can represent the width of such a decay in nonrelativistic approximation as follows:
\begin{equation}
\label{gapp11}
\Gamma^{ZZ}_{\cal{PP}}=\frac{\alpha^3M_Z^2M^3 |\Psi_{\cal P}(0)|^4 \sqrt{\frac{r_3^2}{4}-1}}
{8\pi M_H^2 (r_1^2M_H^2-M_Z^2)(r_2^2M_H^2-M_Z^2)\sin^62\theta_W}\times
\end{equation}
\begin{displaymath}
\left[1+2a_c+2a_b-4a_ca_b-r_3^2+r_3^2a_b+r_3^2a_c-2r_3^2a_ca_b\right]^2,
\end{displaymath}
where $a_c=2|q_c|\sin^2\theta_W$, $a_b=2|q_b|\sin^2\theta_W$.
Numerically, this contribution to the width turns out to be two orders of magnitude 
smaller than the contribution from the quark decay mechanism studied in this work.

As noted above, for simplicity, when constructing decay amplitudes, we took into account 
second-order relativistic corrections both in analytical expressions and in numerical 
calculations, although taking into account corrections of a higher order does not present 
significant difficulties. The main theoretical errors of the obtained results 
for decay widths are connected
with quark masses, coupling constants, fourth-order corrections in relative momenta
$O({\bf p}^4)$, $ O({\bf q}^4)$ and radiative corrections of order $O(\alpha_s)$.
In the case of production amplitudes for the strong coupling constant $\alpha_{c}$ 
we use the leading order approximation from \eqref{26} where the renormalization scale 
$\mu=\frac{m_1}{m_1+m_2}M_H$ and for $\alpha_{b}$ $\mu=\frac{m_2}{m_1+m_2}M_H$.
The total theoretical error of the obtained results can be estimated as
40 percent approximately.

\acknowledgments
The authors are grateful to I.~N.~Belov, A.~V.~Berezhnoy, D.~Ebert, V.~O.~Galkin, 
A.~L.~Kataev, A.~K.~Likhoded for a helpful discussion of various issues related 
to the production of heavy quarkonia.
The work of F.~A.~Martynenko is supported by the Foundation for the
Advancement of Theoretical Physics and Mathematics "BASIS" (grant No. 19-1-5-67-1).

\appendix

\section{Explicit form of the functions $f_i$, $\tilde f_i$, $g_i$, $\tilde g_i$ entering 
the Higgs boson decay widths (30)-(34)}

The functions presented here are expressed in terms of mass ratios:
$\tilde r_1=m_1/M$, $\tilde r_2=m_2/M$, $r_5=M_2/M$, $r_6=M_1/M$, $r_3=M_H/\sqrt{M_1M_2}$.

\begin{equation}
\label{a1}
f_1=- \frac{3}{4}r_6 + \frac{9}{4}r_5 + \frac{3}{4}\tilde r_1\tilde r_2^{-1}r_6 + 
\frac{9}{4}\tilde r_1\tilde r_2^{-1}r_5 + 
3\tilde r_1 - \frac{3}{2}\tilde r_1^2\tilde r_2^{-1} - \frac{3}{4}\rho_1r_5 - 
\frac{3}{4}\rho_1\tilde r_1\tilde r_2^{-1}r_5 + 
\frac{3}{4}\eta_1r_6- \frac{3}{4}\eta_1\tilde r_1\tilde r_2^{-1}r_6,
\end{equation}
\begin{equation}
\label{a2}
g_1= \tilde\omega_{\frac{1}{2}\frac{1}{2}}( \frac{3}{2}\tilde r_1 )+\tilde\omega_{01}
(- \frac{3}{4}\tilde r_1\tilde r_2^{-1}r_6 - \frac{9}{4}\tilde r_1\tilde r_2^{-1}r_5 + 
\frac{3}{2}\tilde r_1^2\tilde r_2^{-1} + 
\frac{3}{4}\tilde r_1^2\tilde r_2^{-1}r_6 + \frac{3}{4}\tilde r_1^2\tilde r_2^{-1}r_5 )
\end{equation}
\begin{displaymath}
+\tilde\omega_{10} ( \frac{3}{4}r_6 - \frac{9}{4}r_5 - \frac{3}{2}\tilde r_1 - 
\frac{3}{4}\tilde r_1r_6 + \frac{3}{4}\tilde r_1r_5 )+\omega_{\frac{1}{2}\frac{1}{2}}( 3\tilde r_1 )
+\omega_{\frac{1}{2}\frac{1}{2}} \tilde\omega_{01}(- \frac{3}{2}\tilde r_1 )+
\omega_{\frac{1}{2}\frac{1}{2}}\tilde\omega_{10} (- \tilde r_1 )+
\end{displaymath}
\begin{displaymath}
\omega_{01}(- \frac{1}{4}r_6 + \frac{3}{4}r_5 - \tilde r_1 + \frac{1}{4}\tilde r_1r_6 - 
\frac{1}{4}\tilde r_1r_5 )+\omega_{01}\tilde\omega_{01}(\frac{1}{4}r_6-\frac{3}{4}r_5 + 
\tilde r_1 -\frac{1}{4}\tilde r_1r_6 + 
\frac{1}{4}\tilde r_1r_5 )
\end{displaymath}
\begin{displaymath}
+\omega_{10}( \frac{1}{4}\tilde r_1\tilde r_2^{-1}r_6 + \frac{3}{4}\tilde r_1\tilde r_2^{-1}r_5 
- \frac{3}{2}\tilde r_1 - \frac{1}{4}\tilde r_1^2\tilde r_2^{-1}r_6 - 
\frac{1}{4}\tilde r_1^2\tilde r_2^{-1}r_5 )+
\end{displaymath}
\begin{displaymath}
\omega_{10}\tilde\omega_{01} ( \frac{1}{2}\tilde r_1 )+\omega_{10}\tilde\omega_{10}
(- \frac{1}{4}\tilde r_1\tilde r_2^{-1}r_6 - \frac{3}{4}\tilde r_1\tilde r_2^{-1}r_5 + 
\frac{3}{2}\tilde r_1
+ \frac{1}{4}\tilde r_1^2\tilde r_2^{-1}r_6 + \frac{1}{4}\tilde r_1^2\tilde r_2^{-1}r_5 ),
\end{displaymath}
\begin{equation}
\label{a3}
f_2=\frac{3}{4}r_6 - \frac{1}{4}r_5 - \frac{3}{4}\tilde r_1\tilde r_2^{-1}r_6 - 
\frac{1}{4}\tilde r_1\tilde r_2^{-1}r_5 - 
\tilde r_1+ \frac{1}{2}\tilde r_1^2\tilde r_2^{-1} + \frac{1}{4}\rho_1r_5 + 
\frac{1}{4}\rho_1\tilde r_1\tilde r_2^{-1}r_5 - 
\frac{1}{4}\eta_1r_6 + \frac{1}{4}\eta_1\tilde r_1\tilde r_2^{-1}r_6,
\end{equation}
\begin{equation}
\label{a4}
g_2=\tilde\omega_{\frac{1}{2}\frac{1}{2}}(- \frac{1}{2}\tilde r_1 )
+ \tilde\omega_{01}( \frac{3}{4}\tilde r_1\tilde r_2^{-1}r_6 + \frac{1}{4}\tilde r_1\tilde r_2^{-1}r_5 - \frac{1}{2}\tilde r_1^2\tilde r_2^{-1}- \frac{1}{4}\tilde r_1^2\tilde r_2^{-1}r_6 - \frac{1}{4}
\tilde r_1^2\tilde r_2^{-1}r_5 )
+ \tilde\omega_{10}(- \frac{3}{4}r_6 + \frac{1}{4}r_5 + 
\end{equation}
\begin{displaymath}
\frac{1}{2}r_1 + \frac{1}{4}\tilde r_1r_6 - 
\frac{1}{4}\tilde r_1r_5)
+ \omega_{\frac{1}{2}\frac{1}{2}} ( \tilde r_1 )
+ \omega_{\frac{1}{2}\frac{1}{2}}\tilde\omega_{01} ( \frac{1}{2}\tilde r_1 )
+ \omega_{\frac{1}{2}\frac{1}{2}}\tilde\omega_{10} (- \tilde r_1 )
+ \omega_{01}(- \frac{3}{4}r_6 + \frac{1}{4}r_5 - \tilde r_1 + \frac{1}{4}\tilde r_1r_6 -
\end{displaymath}
\begin{displaymath}
\frac{1}{4}\tilde r_1r_5 )
+ \omega_{01}\tilde\omega_{01}(\frac{3}{4}r_6 -\frac{1}{4}r_5 + \tilde r_1-
\frac{1}{4}\tilde r_1r_6 + 
\frac{1}{4}\tilde r_1r_5 )
+ \omega_{10}(\frac{3}{4}\tilde r_1\tilde r_2^{-1}r_6 + 
\frac{1}{4}\tilde r_1\tilde r_2^{-1}r_5 + \frac{1}{2}\tilde r_1 - 
\frac{1}{4} \tilde r_1^2\tilde r_2^{-1}r_6 - 
\end{displaymath}
\begin{displaymath}
\frac{1}{4}\tilde r_1^2\tilde r_2^{-1}r_5 )
+ \omega_{10}\tilde\omega_{01}(\frac{1}{2}\tilde r_1 )
+ \omega_{10}\tilde\omega_{10}(-\frac{3}{4}\tilde r_1\tilde r_2^{-1}r_6 - 
\frac{1}{4}\tilde r_1\tilde r_2^{-1}r_5 - 
\frac{1}{2}\tilde r_1+ \frac{1}{4}\tilde r_1^2\tilde r_2^{-1}r_6 + 
\frac{1}{4}\tilde r_1^2\tilde r_2^{-1}r_5 ),
\end{displaymath}
\begin{equation}
\label{a5}
f_3=- \frac{1}{4}r_6 - \frac{1}{4}r_5 - \frac{1}{4}\tilde r_1\tilde r_2^{-1}r_6 + 
\frac{1}{4}\tilde r_1\tilde r_2^{-1}r_5 ),
\end{equation}
\begin{equation}
\label{a6}
g_3=\tilde\omega_{01} ( \frac{1}{4}\tilde r_1\tilde r_2^{-1}r_6 - \frac{1}{4}\tilde r_1\tilde r_2^{-1}r_5 )
+ \tilde\omega_{10} ( \frac{1}{4}r_6 + \frac{1}{4}r_5 )
+ \omega_{01}  (- \frac{1}{12}r_6 - \frac{1}{12}r_5 )
+ \omega_{01}\tilde\omega_{01}( \frac{1}{12}r_6 + \frac{1}{12}r_5 )
\end{equation}
\begin{displaymath}
+ \omega_{10} (-\frac{1}{12}\tilde r_1\tilde r_2^{-1}r_6 + 
\frac{1}{12}\tilde r_1\tilde r_2^{-1}r_5 )
+ \omega_{10}\tilde\omega_{10} ( \frac{1}{12}\tilde r_1\tilde r_2^{-1}r_6 - 
\frac{1}{12}\tilde r_1\tilde r_2^{-1}r_5),
\end{displaymath}
\begin{equation}
\label{a7}
f_4= \frac{3}{4}r_6 - \frac{1}{4}r_5 - \frac{3}{4}\tilde r_1\tilde r_2^{-1}r_6 - 
\frac{1}{4}\tilde r_1\tilde r_2^{-1}r_5 - 
\tilde r_1+ \frac{1}{2}\tilde r_1^2\tilde r_2^{-1} + \frac{1}{4}\rho_1r_5 + 
\frac{1}{4}\rho_1\tilde r_1\tilde r_2^{-1}r_5 - 
\frac{1}{4}\eta_1r_6 + \frac{1}{4}\eta_1\tilde r_1\tilde r_2^{-1}r_6,
\end{equation}
\begin{equation}
\label{a8}
g_4=\tilde\omega_{\frac{1}{2}\frac{1}{2}} (- \frac{1}{2}\tilde r_1 )
+ \tilde\omega_{01}( \frac{3}{4}\tilde r_1\tilde r_2^{-1}r_6 + 
\frac{1}{4}\tilde r_1\tilde r_2^{-1}r_5 - 
\frac{1}{2}\tilde r_1^2\tilde r_2^{-1}- \frac{1}{4}\tilde r_1^2\tilde r_2^{-1}r_6 - 
\frac{1}{4}\tilde r_1^2\tilde r_2^{-1}r_5 )+ 
\end{equation}
\begin{displaymath}
\tilde\omega_{10}(- \frac{3}{4}r_6 + \frac{1}{4}r_5 + \frac{1}{2}\tilde r_1 + 
\frac{1}{4}\tilde r_1r_6 - \frac{1}{4}\tilde r_1r_5)
+ \omega_{\frac{1}{2}\frac{1}{2}} ( \tilde r_1 )
+ \omega_{\frac{1}{2}\frac{1}{2}}\tilde\omega_{01}( \frac{1}{2}\tilde r_1 )
+ \omega_{\frac{1}{2}\frac{1}{2}}\tilde\omega_{10}(  - \tilde r_1 )+
\end{displaymath}
\begin{displaymath}
\omega_{01}(  - \frac{3}{4}r_6 + \frac{1}{4}r_5 - \tilde r_1 + \frac{1}{4}\tilde r_1r_6 - 
\frac{1}{4}\tilde r_1r_5 )
+ \omega_{01}\tilde\omega_{01} ( \frac{3}{4}r_6 - \frac{1}{4}r_5 + \tilde r_1 - 
\frac{1}{4}\tilde r_1r_6 + \frac{1}{4}\tilde r_1r_5 )+
\end{displaymath}
\begin{displaymath}
\omega_{10}( \frac{3}{4}\tilde r_1\tilde r_2^{-1}r_6 + \frac{1}{4}\tilde r_1\tilde r_2^{-1}r_5 + \frac{1}{2}r_1 - 
\frac{1}{4}\tilde r_1^2\tilde r_2^{-1}r_6 - \frac{1}{4}\tilde r_1^2\tilde r_2^{-1}r_5 )
+ \omega_{10}\tilde\omega_{01} ( \frac{1}{2}\tilde r_1 )
\end{displaymath}
\begin{displaymath}
+ \omega_{10}\tilde\omega_{10}(-\frac{3}{4}\tilde r_1\tilde r_2^{-1}r_6 - 
\frac{1}{4}\tilde r_1\tilde r_2^{-1}r_5 - 
\frac{1}{2}\tilde r_1+ \frac{1}{4}\tilde r_1^2\tilde r_2^{-1}r_6 + 
\frac{1}{4}\tilde r_1^2\tilde r_2^{-1}r_5 ),
\end{displaymath}
\begin{equation}
\label{a9}
f_5=- \frac{3}{2}r_6 + \frac{1}{2}r_5 - \frac{3}{2}\tilde r_1\tilde r_2^{-1}r_6 - 
\frac{1}{2}\tilde r_1\tilde r_2^{-1}r_5 + 
2r_1 + \tilde r_1^2\tilde r_2^{-1} - \frac{1}{2}\rho_1r_5 + 
\frac{1}{2}\rho_1\tilde r_1\tilde r_2^{-1}r_5 + 
\frac{1}{2}\eta_1r_6 + \frac{1}{2}\eta_1\tilde r_1\tilde r_2^{-1}r_6,
\end{equation}
\begin{equation}
\label{a10}
g_5= \tilde\omega_{\frac{1}{2}\frac{1}{2}} (- \tilde r_1 )
+ \tilde\omega_{01}( \frac{3}{2}\tilde r_1\tilde r_2^{-1}r_6 + \frac{1}{2}\tilde r_1
\tilde r_2^{-1}r_5 - 
\tilde r_1^2\tilde r_2^{-1} - 
\frac{1}{2}\tilde r_1^2\tilde r_2^{-1}r_6 - \frac{1}{2}\tilde r_1^2\tilde r_2^{-1}r_5 )+
\end{equation}
\begin{displaymath}
\tilde\omega_{10} ( \frac{3}{2}r_6 - \frac{1}{2}r_5 - \tilde r_1 - \frac{1}{2}\tilde r_1r_6 + 
\frac{1}{2}\tilde r_1r_5 )
+ \omega_{\frac{1}{2}\frac{1}{2}} (  - 4\tilde r_1 )
+ \omega_{\frac{1}{2}\frac{1}{2}}\tilde\omega_{01}( \tilde r_1 )
+ \omega_{\frac{1}{2}\frac{1}{2}}\tilde\omega_{10}( 2\tilde r_1 )+
\end{displaymath}
\begin{displaymath}
\omega_{01} (  - \frac{3}{2}r_6 + \frac{1}{2}r_5 - 2\tilde r_1 + \frac{1}{2}\tilde r_1r_6 - 
\frac{1}{2}\tilde r_1r_5 )
+ \omega_{01} \tilde\omega_{01} ( \frac{3}{2}r_6 - \frac{1}{2}r_5 + 2\tilde r_1 - 
\frac{1}{2}\tilde r_1r_6 + \frac{1}{2}\tilde r_1 r_5 )+
\end{displaymath}
\begin{displaymath}
\omega_{10}  (  - \frac{3}{2}\tilde r_1\tilde r_2^{-1}r_6 - \frac{1}{2}\tilde r_1\tilde r_2^{-1}r_5 - 
r_1 + \frac{1}{2}\tilde r_1^2\tilde r_2^{-1}r_6 + \frac{1}{2}\tilde r_1^2\tilde r_2^{-1}r_5 )
+ \omega_{10}\tilde\omega_{01} ( \tilde r_1 )+
\end{displaymath}
\begin{displaymath}
\omega_{10}\tilde\omega_{10}( \frac{3}{2}\tilde r_1\tilde r_2^{-1}r_6 + 
\frac{1}{2}\tilde r_1\tilde r_2^{-1}r_5 + \tilde r_1 - \frac{1}{2}\tilde r_1^2\tilde r_2^{-1}r_6 - 
\frac{1}{2}\tilde r_1^2\tilde r_2^{-1}r_5 ).
\end{displaymath}

The functions $\tilde f_i$, $\tilde g_i$ can be obtained from $f_i$, $g_i$ as a result 
of the replacement $\tilde r_1\leftrightarrow \tilde r_2$, 
$\tilde\omega_{ij}\leftrightarrow \tilde\omega_{ji}$.

\section{The construction of an effective Hamiltonian}

Using potential \eqref{eq:breit}, we construct an effective model of the quark interaction of the 
Schr\"odinger-type and calculate the relativistic parameters included in the decay widths.
The first step in the construction of the model is related to the rationalization of the kinetic 
energy operator as follows:
\begin{equation}
\label{b1}
T=\sqrt{{\bf p}^2+m_1^2}-m_1+\sqrt{{\bf p}^2+m_2^2}-m_2=
\frac{{\bf p}^2}{\sqrt{{\bf p}^2+m_1^2}+m_1}+\frac{{\bf p}^2}{\sqrt{{\bf p}^2+m_2^2}+m_2}
\approx 
\end{equation}
\begin{displaymath}
\frac{{\bf p}^2}{E_1+m_1}+\frac{{\bf p}^2}{E_2+m_2}=
\frac{{\bf p}^2}{2\tilde m_1}+\frac{{\bf p}^2}{2\tilde m_2}=\frac{{\bf p}^2}{2\tilde \mu},
\end{displaymath}
where we change relativistic particle energies $\varepsilon_{1,2}({\bf p})$ by
their effective values $E_{1,2}$ so that $M_{B_c}=E_1+E_2$
or what is the same ${\bf p}^2\to {\bf p}^2_{eff}$. 
The effective quark masses
$\tilde m_{1,2}$ are used further in the program for the numerical solution of the 
Schr\"odinger equation.
${\bf p}^2_{eff}$ should be considered as a new parameter which
effectively accounts for relativistic corrections.

Another part of the relativistic corrections in the Breit Hamiltonian is determined 
by the term:
\begin{equation}
\label{b2}
\Delta\tilde U=-\frac{2\alpha_s}{3m_1m_2r}\left[{\bf p}^2-\frac{d^2}{dr^2}\right].
\end{equation}

In order to replace it by the effective term containing the power-like potentials, 
we use the approximate bound state wave functions which can be written for S- and 
P-wave states as follows:
\begin{equation}
\Psi_{\cal S}({\bf r})=\frac{\beta^{3/2}}{\pi^{3/4}}e^{-\frac{1}{2}\beta^2r^2},~~
\Psi_{\cal P}({\bf r})=\sqrt{\frac{8}{3}}\frac{\beta^{3/2}}{\pi^{3/4}}\beta
re^{-\frac{1}{2}\beta^2r^2}Y_{1m}(\theta,\phi).
\end{equation}
The parameter $\beta$ entering here is chosen in such a way that the calculated 
value of $R_S(0)$ and $R'_P(0)$ is obtained.
The operator ${\bf p}^2$ is changed by its nonrelativistic expression: 
${\bf p}^2=2\mu \left(E_{nr}+\frac{4\alpha_s}{3r}-Ar-B\right)$, where nonrelativistic
energy $E_{nr}$ is obtained by the numerical solution of the Schr\"odinger equation in
the nonrelativistic approximation.

In the case of S-wave states it is necessary to change the $\delta$-like term of the potential
by known smeared $\delta$-function of the Gaussian form:
\begin{equation}
\label{b3}
\tilde\delta({\bf r})=\frac{b^3}{\pi^{3/2}}e^{-b^2r^2}
\end{equation}
with the additional parameter $b$ which defines the hyperfine splitting in the $(b\bar c)$ 
system. 

The second step in constructing an effective model of quark interaction is connected 
with the angular averaging of the spin-orbit and spin-spin terms. This averaging is 
performed separately for the considered S- and P-states \cite{tom2}.
In this case, we use a basis transformation of the following form:
\begin{equation}
\label{b4}
\Psi_{SLFM_F}=\sum_{J}(-1)^{s_1+s_2+L+F}\sqrt{(2S+1)(2J+1)}
\Biggl\{
\begin{array}{ccc}
 L&~s_1&~J\\
s_2&~F&~S
\end{array}
\Biggr\}\Psi_{Js_2FM_F},
\end{equation}
where ${\bf J}={\bf L}+{\bf s_1}$ is the total nomentum of quark 1, ${\bf S}={\bf s}_1+{\bf s}_2$,
${\bf F}={\bf S}+{\bf L}={\bf J}+{\bf s}_2$.
For diagonal matrix elements we have the expression:
\begin{equation}
\label{b5}
<Js_2FM_F|[({\bf L}{\bf s}_2)+3({\bf s}_1{\bf n})({\bf s}_2{\bf n})-({\bf s}_1{\bf s}_2)]|Js_2FM_F>=
\frac{L(L+1)}{J(J+1)}[F(F+1)-J(J+1)-s_2(s_2+1)].
\end{equation}
The matrix elements of the spin-orbit $T_1={\bf L}{\bf s}_2$ and spin-spin 
interactions $T_2=3({\bf s}_1{\bf n})({\bf s}_2{\bf n})-({\bf s}_1{\bf s}_2)$
off-diagonal in $J$ with $F=1$ are determined as follows:
\begin{equation}
\label{b6}
<J's_2FM_F|T_1|Js_2FM_F>=
(-1)^{-J-F-s_2+L+3/2+J'}\sqrt{(2J'+1)(2J+1)}\times
\end{equation}
\begin{displaymath}
\sqrt{(2s_2+1)(s_2+1)s_2(2L+1)(L+1)L}
\Biggl\{
\begin{array}{ccc}
 J&~s_2&~F\\
s_2&~J'&~1
\end{array}
\Biggr\}
\Biggl\{
\begin{array}{ccc}
 L&~J'&~1/2\\
J&~L&~1
\end{array}
\Biggr\}.
\end{displaymath}

As a result $<J's_2FM_F|T_1|Js_2FM_F>=2<J's_2FM_F|T_2|Js_2FM_F>=-\frac{\sqrt{2}}{3}$.
After angular averaging an effective Hamiltonians are obtained describing the states 
$^3S_1$, $^1S_0$, $^3P_J$, $^1P_1$.

Finally, the third final step is related to the numerical solution of the Schr\"odinger 
equation with obtained effective Hamiltonians.
We use a calculation program in the system Mathematica \cite{LS}.
The obtained numerical results are presented in Table~\ref{tb1}.
They are in good agreement with previous calculations \cite{glko,rqm1,rqm2}.

\end{document}